\documentclass[12pt,letterpaper]{iopart}
\usepackage[bookmarks,pdfhighlight=/O,pdfstartview=FitH]{hyperref}
\usepackage{iopams,mathrsfs,graphicx,epstopdf}
\usepackage{color}

\newcommand{\Lag}{\mathscr{L}}

\newcommand{\im}{\mathrm{Im}}
\newcommand{\dd}{\mathrm{d}}

\begin{document}
\title{The $\Delta$(1232) at RHIC}
\author{Hendrik van Hees\footnote[1]{presenting author} and Ralf Rapp}
\address{Cyclotron Institute and Physics Department, Texas A\&M 
University, College Station, Texas 77843-3366}
\ead{hees@comp.tamu.edu, rapp@comp.tamu.edu}
\begin{abstract}
  We investigate properties of the $\Delta$(1232) and nucleon spectral
  functions at finite temperature and baryon density within a hadronic
  model. The medium modifications of the $\Delta$ consist of a
  renormalization of its pion-nucleon cloud and resonant $\pi\Delta$
  scattering. Underlying coupling constants and form factors are determined
  by the elastic $\pi N$ scattering phase shift in the isobar channel, as
  well as empirical partial decay widths of excited baryon resonances.
  For cold nuclear matter the model provides reasonable agreement with
  photoabsorption data on nuclei in the $\Delta$-resonance region.  In hot
  hadronic matter typical for late stages of central $Au$-$Au$ collisions
  at RHIC we find the $\Delta$-spectral function to be broadened by
  $\sim$65~MeV together with a slight upward mass shift of 5-10 MeV, in
  qualitative agreement with preliminary data from the STAR collaboration.
\end{abstract}

\section{Introduction}

At low energies, the main features of Quantum Chromodynamics (QCD) are
confinement and the spontaneous breaking of chiral symmetry.  The former
implies that we only observe hadrons (rather than quarks and gluons), while
the latter is believed to govern the (low-lying) hadron-mass spectrum.
Lattice-QCD calculations predict a phase transition from nuclear/hadronic
matter to a deconfined, chirally symmetric state~\cite{kar01} at
temperatures $T$$\simeq$150-200~MeV, dictating a major reshaping of the
hadronic spectrum in terms of degenerate chiral partners.  The observation
of such medium modifications is therefore an important objective in
relativistic heavy-ion collision experiments.

Large theoretical efforts have been devoted to evaluate in-medium
properties of vector mesons which are accessible experimentally through
dilepton invariant-mass spectra~\cite{rw99}. In most of these studies,
baryon-driven effects are essential to account for the dilepton enhancement
observed in $Pb$-$Au$ collisions at the SPS below the free $\rho$
mass~\cite{ceres98,ceres03}.  Thus, changes in the baryon properties
themselves deserve further investigation. In addition, recent measurements
of $\pi N$ invariant-mass spectra in nuclear
collisions~\cite{hjo97,pelte97,fach04a} may open a more direct
window on modifications of the $\Delta$(1232).

To date, in-medium properties of the $\Delta$ have mostly been assessed in
cold nuclear matter~\cite{hori80,os87,ew88,mig90,xia94,lutz03}, with few
exceptions~\cite{ko89,korpa95}.  In this article we will discuss properties
of the nucleon and the $\Delta$(1232) in a hot and dense
medium~\cite{hr04}, employing a finite-temperature field theory framework
based on hadronic interactions. Both direct interactions of the $\Delta$
with thermal pions as well as modifications of its free $\pi N$ self-energy
(incuding vertex corrections) will be accounted for.

The article is organised as follows. In Sec.~\ref{sect.2} we introduce the
hadronic Lagrangian and outline how its parameters are determined using
scattering and decay data in vacuum. In Sec.~\ref{sect.3} we compute
in-medium self-energies for nucleon and $\Delta$. In Sec.~\ref{sect.4} we
first check our model against photoabsorption cross sections on the nucleon
and nuclei, followed by a discussion of the spectral functions under
conditions expected to occur in high-energy heavy-ion collisions.  We close
with a summary and outlook.

\section{Hadronic Interaction Lagrangians in Vacuum}
\label{sect.2}
The basic element of our analysis are 3-point interaction vertices
involving a pion and two baryons, $\pi B_1 B_2$. Baryon fields are treated
using relativistic kinematics, $E_B^2(\bi{p})=m_B^2+\bi{p}^2$, but
neglecting anti-particle contributions and restricting Rarita-Schwinger
spinors to their non-relativistic spin-3/2 components. Pions are treated
fully relativisticly ($\omega_{\pi}^2(\bi{k})=m_{\pi}^2+\bi{k}^2$). The
resulting interaction Lagrangians are thus of the usual nonrelativistic
form involving (iso-) spin-1/2 Pauli matrices, 1/2 to 3/2 transition
operators, as well as spin-3/2
matrices~\cite{os87,cub90,rubw98,ubw99,novr01}, see Ref.~\cite{hr04} for
explicit expressions.  To simulate finite-size effects we employed hadronic
form factors with a uniform cutoff parameter $\Lambda_{\pi B_1
  B_2}=500$~MeV (except for $\pi NN$ and $\pi N\Delta$ vertices).

The imaginary part of the vacuum self-energy for the $\Delta$$\rightarrow$$N\pi$ decay
takes the form 
\begin{equation}
\label{3}
\im \Sigma_{\Delta}^{(N\pi)}(M) = -\frac{f_{\pi N \Delta}^2}{12 m_{\pi}^2 \pi}
\frac{m_N k_{\mathrm{cm}}^3}{M} F^2(k_{\mathrm{cm}}) \Theta(M-m_N-m_{\pi})
\end{equation}
with $k_{\mathrm{cm}}$ the
center-of-mass decay momentum (an extra factor $m_N/E_N(k_{\mathrm{cm}})$
has been introduced in Eq. (\ref{3}) to restore Lorentz-invariance), and
the real part is determined via a dispersion relation. With a bare mass of
$m_{\Delta}^{(0)}$=1302~MeV, a form-factor cutoff $\Lambda_{\pi N
  \Delta}$=290~MeV and a coupling constant $f_{\pi N \Delta}=3.2$ we obtain
a satisfactory fit to the experimental $\delta_{33}$-phase
shift~\cite{mon80,korpa95,wfn98}.

To account for resonant interactions of the $\Delta$ with pions we
identified the relevant excited baryons via their decay branchings
$B$$\to$$\pi\Delta$. The pertinent coupling constants have been determined
assuming the lowest partial wave to be dominant (unless otherwise
specified)~\cite{pdb02}, using pole masses and (total) widths of the
resonances. The same procedure has been adopted for resonant $\pi N$
interactions (which are used to evaluate finite-temperature effects on the
nucleon).  The resonances included are $N$(1440), $N$(1520), $N$(1535),
$\Delta$(1600), $\Delta$(1620) and $\Delta$(1700).  The total widths
figuring into the resonance propagators have been obtained by scaling up
the partial $\pi N$ and $\pi\Delta$ channels, and vacuum renormalizations
of the masses have been neglected.

Finally, the evaluation of the photoabsorption cross section requires a
$\gamma N \Delta$ vertex for which we employ the magnetic
coupling~\cite{ew88}
\begin{equation}
\label{4}
\Lag_{\gamma N \Delta}=-\frac{f_{\gamma N \Delta}}{4 \pi m_{\pi}}
\psi_N^{\dagger} (\bi{S}^{\dagger} \times \nabla) \bi{A} T_3^{\dagger}
\Psi_{\Delta} + \mathrm{h.c.}
\end{equation}

\section{Self-energies at Finite Temperature and Density}
\label{sect.3}
The interactions described in the previous section are now used to evaluate
the $\Delta$ self-energy in hot hadronic matter.  The first modification
concerns the $\pi N$ loop which we obtain within the Matsubara formalism as
\begin{eqnarray}
\label{sgdnpi}
\Sigma_{\Delta}^{(N\pi)}(p) = & \frac{f_{\pi N \Delta}^2}{3 m_{\pi}^2} \int
 \frac{\dd^4 l}{(2 \pi)^4} 
\frac{m_N}{E_N(\bi{l})} \bi{k}^2 F_{\pi}^2(|\bi{k}|) \\ 
& \times \lbrace [\Theta(k_0)+\sigma(k_0) f^{\pi}(|k_0|)] {A}_{\pi}(k) G_N(l)
-f^N(l_0) A_N(l) G_{\pi}(k) \rbrace \ , \nonumber 
\end{eqnarray}
where $k=p-l$ is the pion 4-momentum. 
The thermal distributions are defined by
$f^N(l_0)$=$f^{\mathrm{fermi}}(l_0-\mu_N,T)$ and
$f^{\pi}(|k_0|)$=$f^{\mathrm{bose}}(|k_0|,T) \exp(\mu_{\pi}/T)$, with
$f^{\mathrm{fermi}}$ and $f^{\mathrm{bose}}$ the Fermi and Bose functions,
respectively. For simplicity, finite pion-chemical potentials,
$\mu_{\pi}$$>$0, are treated in the Boltzmann limit to avoid Bose
singularities in the presence of broad pion spectral functions (a more
detailed discussion of this point will be given elsewhere).  In
Eq.~(\ref{sgdnpi}) positive energies $k_0$$>$0 correspond to outgoing
pions, i.e., $\Delta$$\rightarrow$$\pi N$ decays, while $k_0$$<$0 accounts
for scattering with (incoming) pions from the heat bath.

\begin{figure}[t]
\begin{center}
\parbox{0.3\linewidth}{\includegraphics[width=\linewidth]{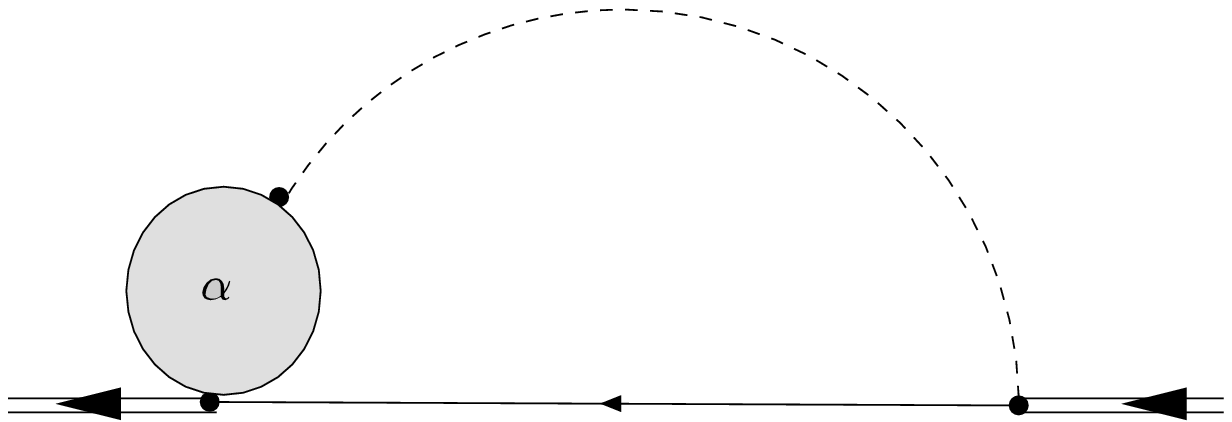}}\hspace*{2mm}
\raisebox{-0.41mm}{\parbox{0.3\linewidth}{\includegraphics[width=\linewidth]{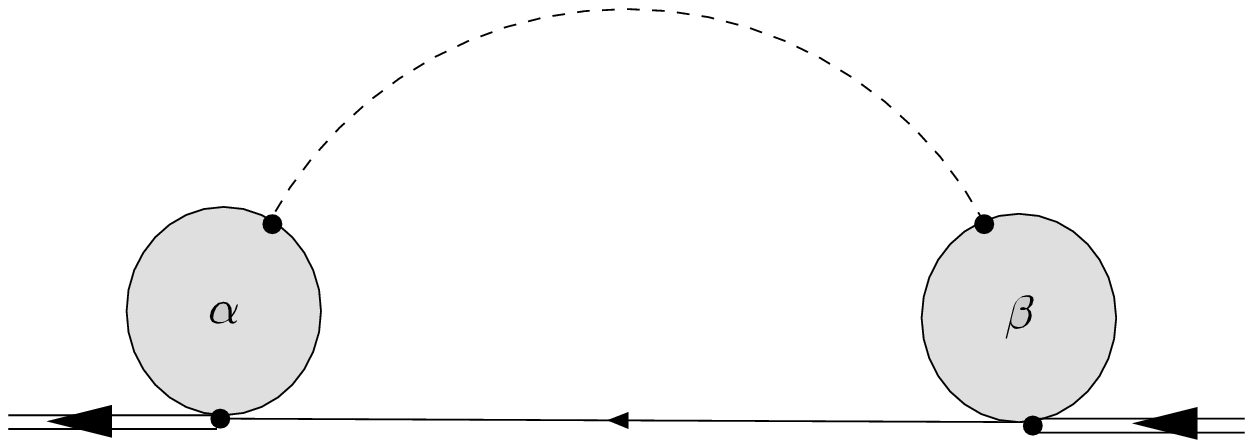}}}\hspace*{2mm}
\raisebox{-7.0mm}{\parbox{0.25\linewidth}{\includegraphics[width=\linewidth]{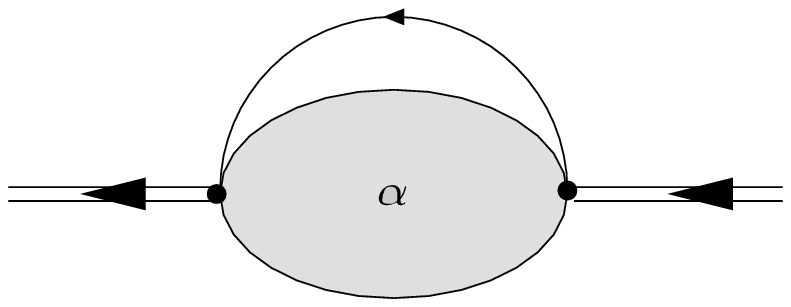}}}
\end{center}
\caption{Diagrammatic representation of $\pi N\Delta$ vertex corrections
  (dashed lines: pion, solid lines: nucleon, double solid lines: 
   $\Delta(1232)$); a bubble with label $\alpha$
  corresponds to a Lindhard function $\Pi_{\alpha}$ ($\alpha \in
  \{1,2\}$) attached to baryon lines with pertinent Migdal
  parameters, i.e., $g_{12}'$ for $\alpha$$=$1 and $g_{22}'$ for
  $\alpha$$=$2.}
\label{fig.vcorr}
\end{figure}

The key quantities in Eq.~(\ref{sgdnpi}) are the in-medium pion and nucleon
propagators, $G_\pi$ and $G_N$, and pertinent spectral functions
$A_N$=$-2\im G_N$ and $A_{\pi}$=$-2 \im G_{\pi}$.  The modifications of the
pion propagator are implemetend via a self-energy, arising from two parts:
(i) interactions with thermal pions modeled by a four-point interaction in
second order (``sunset diagram")~\cite{vHK2001-Ren-II}, with a coupling
constant adjusted to qualitatively reproduce the results of more elaborate
$\pi\pi$ interactions in $s$, $p$, and $d$-wave~\cite{rw95}; (ii)
interactions with baryons via $p$-wave nucleon- and $\Delta$-hole
excitations at finite temperature, described by standard Lindhard
functions, supplemented by short-range correlations encoded in Migdal
parameters~\cite{mig78} (our default values are $g'_{NN}$=0.8,
$g'_{N\Delta}$=$g'_{\Delta\Delta}$=0.33).  These excitations induce a
softening of the pion-dispersion relation which can even lead to a (near)
vanishing of the pion group velocity at finite momentum, inducing an
artificial threshold enhancement in the $\Delta$ self-energy~\cite{ko89}.
This feature is remedied by accounting for appropriate vertex corrections,
which in the case of $\rho$$\to$$\pi\pi$ decays are required to maintain a
conserved vector current in the medium~\cite{cs93,hfn93}. Here we apply the
same technique to the $\pi N\Delta$ vertex, cf. Fig.~\ref{fig.vcorr}.
\begin{figure}[t]
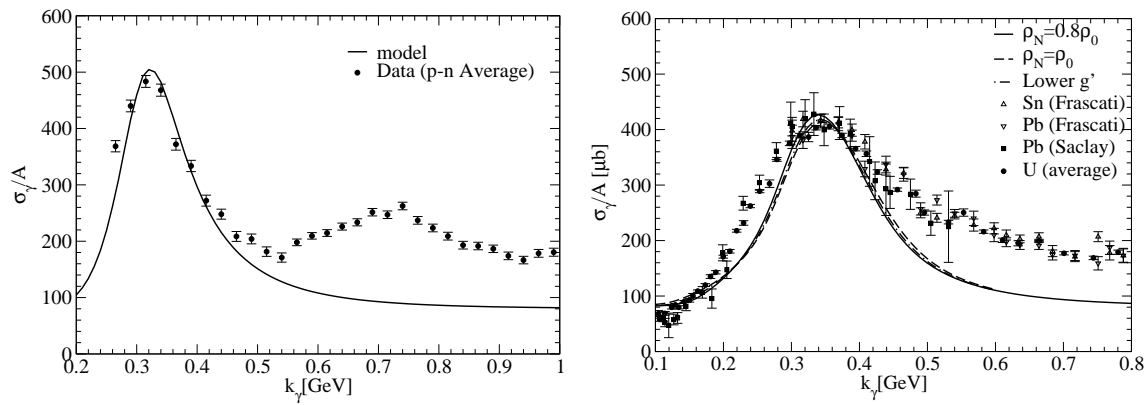

\begin{minipage}{0.47\textwidth}
\includegraphics[width=\textwidth]{photo-absorption-nucl}
\end{minipage}\hspace*{2mm}
\begin{minipage}{0.47\textwidth}
\includegraphics[width=\textwidth]{photo-absorption-proc}
\end{minipage}
  \caption{Photoabsorption cross sections on nucleons (left 
    panel, data from \cite{lep78}) and nuclei (right panel; data from
    \cite{ahr84,ahr85,fro92,fro94,bian93,bian96}).\label{fig_photo}}
\end{figure}
The nucleon self-energy is calculated in terms of resonant interactions
with thermal pions, at the same level of approximation as the 
pion Lindhard functions (i.e., neglecting off energy-shell dependencies 
in the spectral functions of the excited baryons).

The second contribution to the in-medium $\Delta$ self-energy consists of 
resonant $\pi\Delta\to B$ interactions, corresponding to the    
finite-temperature part of $\pi B$ loops. The resulting self-energy 
expressions are similar to Eq.~(\ref{sgdnpi}) but with only
the scattering part ($k_0<0$) retained (note that this is consistent
with our description of the $\Delta$(1232) in vacuum where $\pi B$ 
loops are not included).

\section{In-medium Spectral Properties of the $\Delta$}
\label{sect.4}

\subsection{Photoabsorption on Nucleons and Nuclei}
\label{sec_photo}
Valuable constraints on the $\Delta$ spectral function in cold
nuclear matter can be obtained from photoabsorption cross sections 
on nuclei. To leading order in $\alpha_{\rm em}$, 
the latter can be related to the photon self-energy (electromagnetic 
current correlator), $\Pi_\gamma$, by~\cite{rubw98}
\begin{equation}
\label{photoabs}
\frac{\sigma_{\gamma A}^{\mathrm{abs}}}{A}=\frac{4 \pi \alpha}{k}
\frac{1}{\varrho_N} \frac{1}{2} \im \Pi_{\gamma}(k_0=k), \quad 
\Pi_{\gamma}=\frac{1}{2} g_{\mu \nu} \Pi^{\mu \nu}, 
\label{sig_gam}
\end{equation}
where $k_0=k$ denotes the photon energy (momentum). With the vertex of
Eq.~(\ref{4}) we evaluate the $\gamma$-induced $\Delta$-hole loop using our
full $\Delta$ propagator.  The cross section for a nucleon target follows
from the low-density limit ($\varrho_N \rightarrow 0$) of
Eq.~(\ref{sig_gam}) involving the free $\Delta$-spectral function, which we
use to fix the coupling constant and form-factor cut-off of the $\gamma
N\Delta$ vertex at $f_{\gamma N \Delta}$=0.653 and $\Lambda_{\gamma N
  \Delta}$=400 MeV, respectively, cf. left panel of Fig.~\ref{fig_photo}
(we have also included an estimate of the nonresonant background of
$80\mu$b~\cite{rubw98}).
\begin{figure}[t]
\begin{minipage}{0.47\textwidth}
\includegraphics[width=\textwidth]{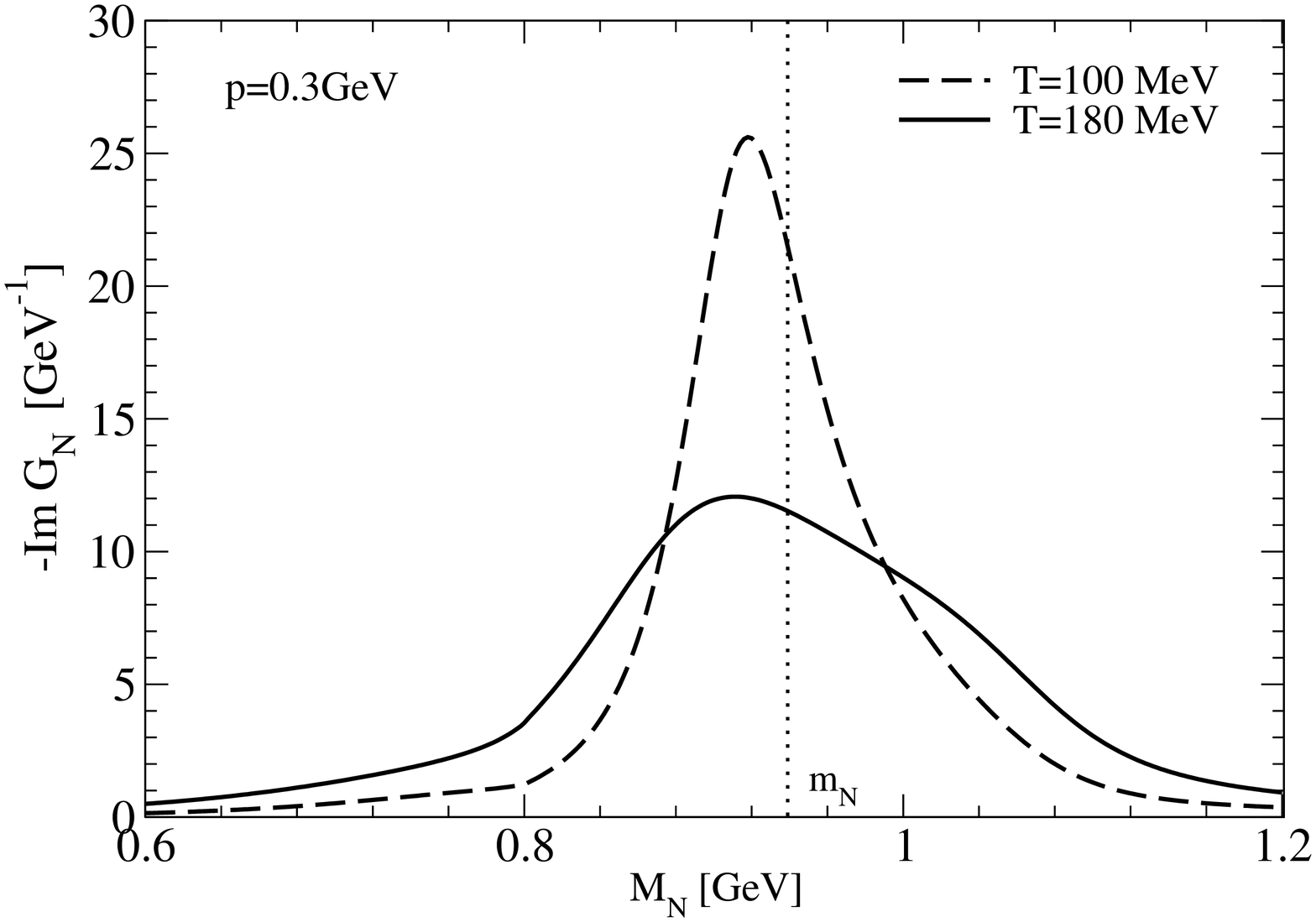}
\end{minipage}\hspace*{2mm}
\begin{minipage}{0.47\textwidth}
\includegraphics[width=\textwidth]{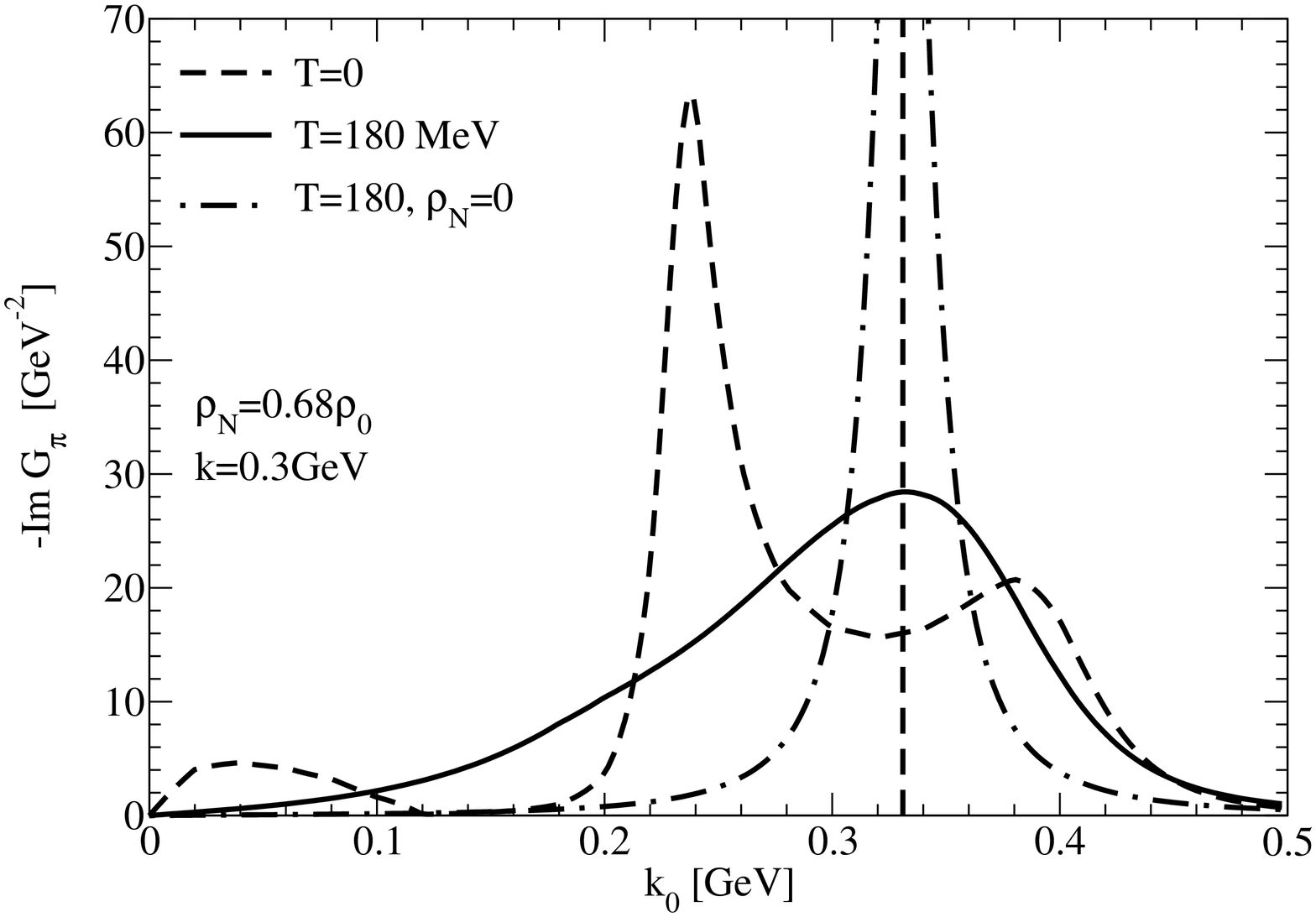}
\end{minipage}
\caption{Left panel: nucleon spectral function at RHIC
  (solid line $T$=180 MeV, $\varrho_N$=0.68$\varrho_0$; 
dashed line: $T$=100 MeV, $\varrho_N$=0.12 $\varrho_0$). 
Right panel: pion spectral function for cold
nuclear matter (dashed line: $T$=0, $\varrho_N$=0.68$\varrho_0$) and at 
RHIC (solid line: $T$=180 MeV, $\varrho_N$=0.68$\varrho_0$); the dash-dotted
line corresponds to switching off baryonic effects leaving only
the 4-point interactions with thermal pions.}
\label{fig_nuc-pi}
\end{figure}
Taking an average nuclear density of $0.8\varrho_0$, our prediction for
nuclei follows without further assumption, cf. right panel of
Fig.~\ref{fig_photo}. The sensitivity to changes in the Migdal parameters
or the nuclear density is very moderate.  Given our rather simple approach
for the cross section, the agreement with data is fair.  The discrepancies
at low energy (which seem to be present already for the nucleon)
could be due to interference with the background, collective
effects involving direct $N$$N^{-1}$-excitations, or transverse contributions
with in-medium $\rho$ mesons in the vertex corrections of the $\Delta$
decay. At higher energies, further resonances in the photon self-energy
need to be included.

\subsection{Hot Hadronic Matter}
Let us finally turn to the results for hot hadronic matter.  In heavy-ion
collisions one expects a hierarchy of chemical freeze-out (determining the
ratios of stable hadrons) and thermal freeze-out (where elastic
rescattering ceases).  The former is characterized by a temperature
$T_{\mathrm{chem}}$ and a common baryon chemical potential $\mu_B$. Thermal
freezeout occurs at lower $T_{\mathrm{fo}}$$\simeq$100~MeV, which requires
the build-up of additional chemical potentials for pions, kaons,
etc.~\cite{rap02}, to conserve the observed hadron ratios, including
relative chemical equilibrium for elastic processes, e.g. $\pi N$$\leftrightarrow$$\Delta$ implying $\mu_{\Delta}$=$\mu_N+\mu_{\pi}$.

\begin{figure}[t]
\begin{center}
\begin{minipage}{0.47\textwidth}
\includegraphics[width=\textwidth]{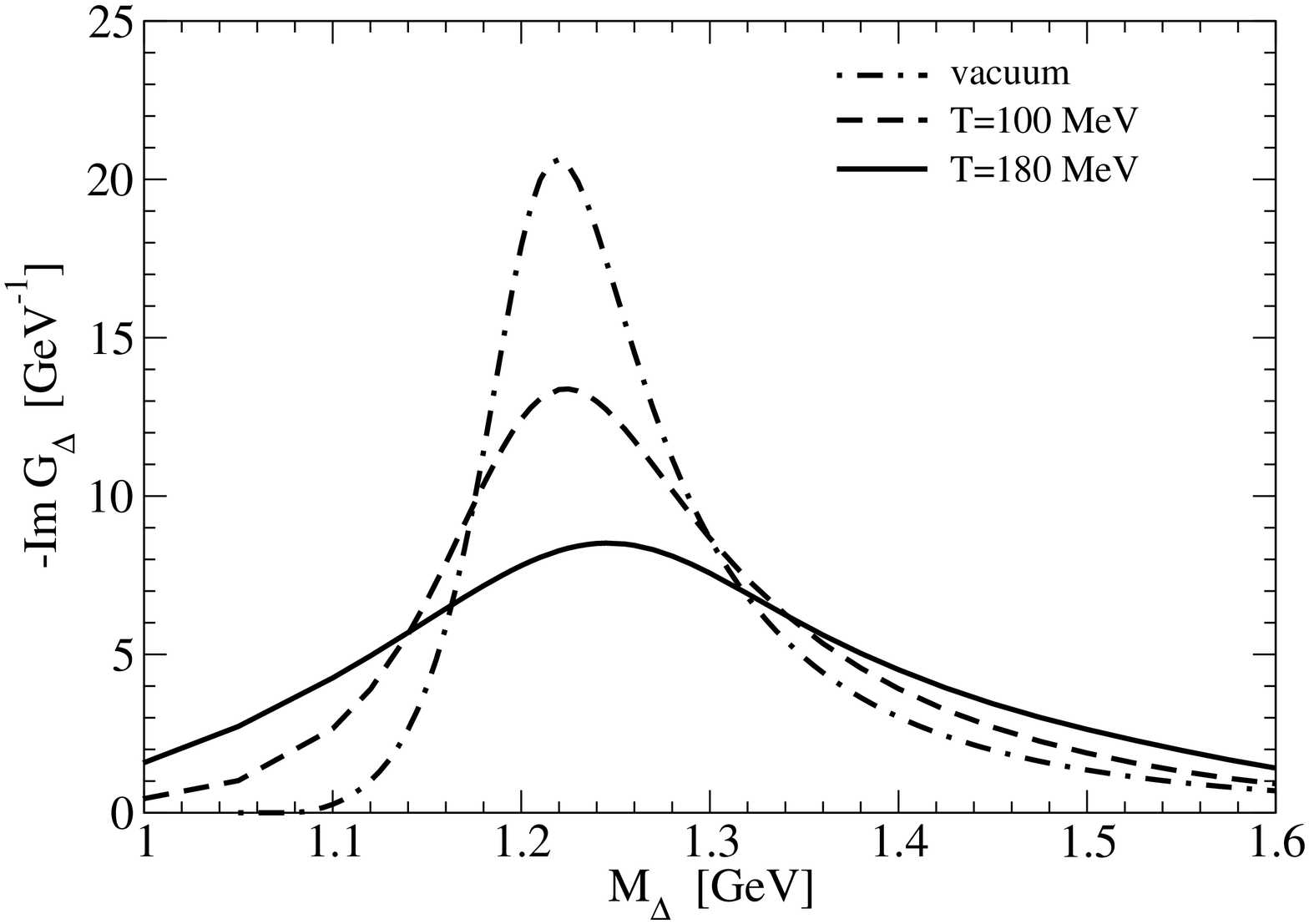}
\end{minipage}
\begin{minipage}{0.47\textwidth}
\includegraphics[width=\textwidth]{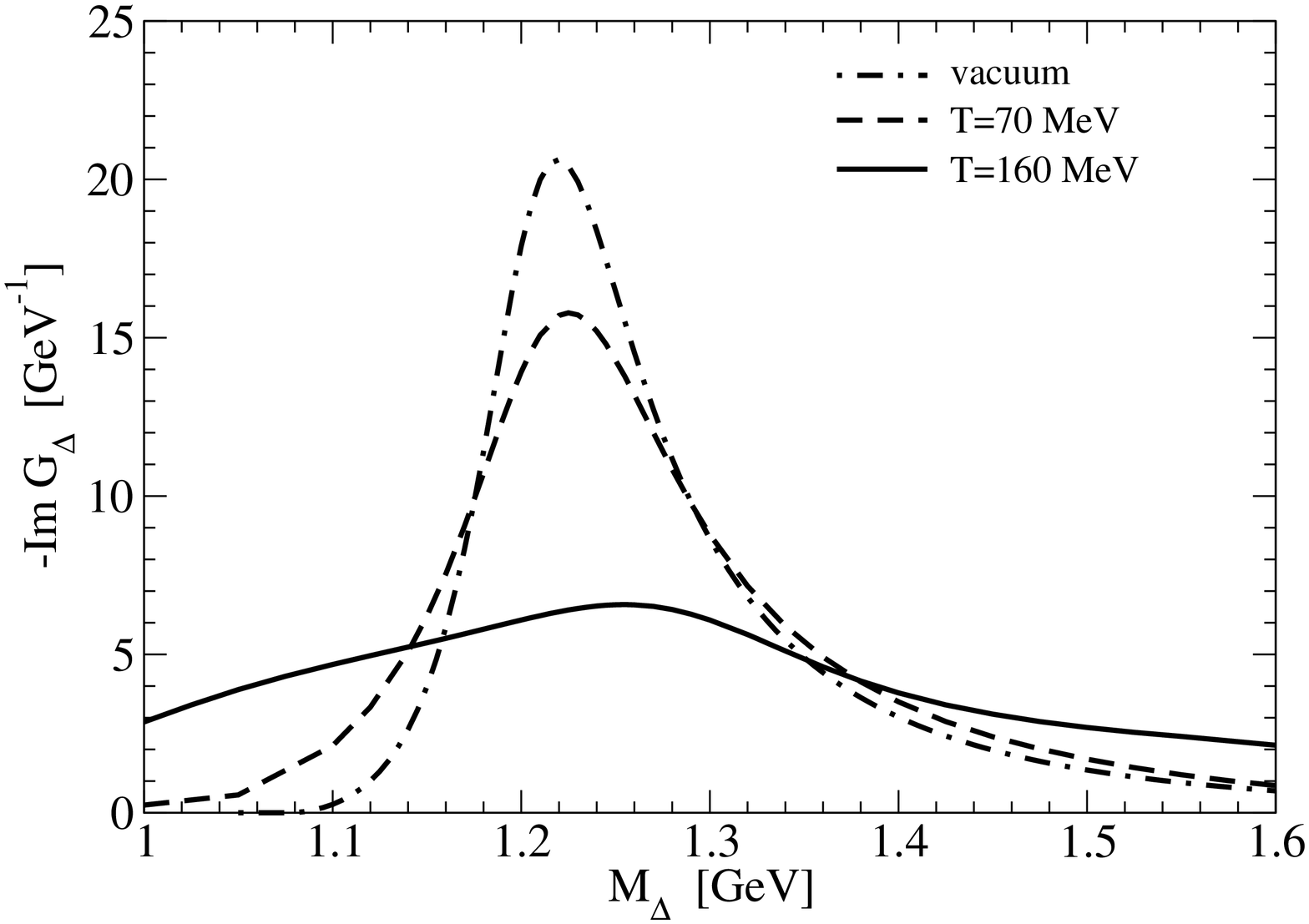}
\end{minipage}
\end{center}
\caption{In-medium $\Delta$(1232) spectral functions in heavy-ion collisions
  compared to free space (dash-dotted lines);
  left panel: RHIC;  
  dashed line: $T$=100~MeV, $\varrho_N$$=$0.12$\varrho_0$
  ($\mu_N$=531~MeV), $\mu_{\pi}$$=$96~MeV; solid line: $T$$=$180~MeV,
  $\varrho_N$$=$0.68$\varrho_0$ ($\mu_N$=333~MeV),
  $\mu_{\pi}$$=$0. Right: future GSI facility;  dashed line: $T$$=$70~MeV,
  $\varrho_N$$=$0.19$\varrho_0$ ($\mu_N$$=$727~MeV), $\mu_{\pi}$$=$105~MeV;
  solid line: $T$$=$160~MeV, $\varrho_N$$=$1.80$\varrho_0$
  ($\mu_N$$=$593~MeV), $\mu_{\pi}$$=$0.}
\label{fig_del}
\end{figure}

Under RHIC conditions the nucleon spectral function exhibits an appreciable
broadening and a moderate downward mass shift (left panel of
Fig.~\ref{fig_nuc-pi}) due to resonant scattering off thermal pions. The
pion spectral function (right panel of Fig.~\ref{fig_nuc-pi}) is strongly
broadened mostly due to scattering off baryons, with little mass shift.
Thermal motion completely washes out the multi-level structure visible at
zero temperature (dashed line).  Also for the $\Delta$ spectral function
(left panel in Fig.~\ref{fig_del}) the main effect is a broadening with a
slight repulsive mass shift.  Half of the increase of the in-medium width
is due to baryon-resonance excitations (slightly enhanced due to in-medium
pion propagators), adding to the contribution of the $\pi N$ loop. In the
real part, however, the predominantly repulsive contributions from baryon
resonances are counterbalanced by net attraction in the $\pi N$ loop
(mostly due to the pion-Bose factor). At thermal freeze-out we find a peak
position at about $M$$\simeq$1.226~GeV and a width $\Gamma$$\simeq$177~MeV,
to be compared to the corresponding vacuum values of $M$$\simeq$1.219~GeV
and $\Gamma$$\simeq$110~MeV, in qualitative agreement with pre\-liminary
data from STAR~\cite{fach04a}.  For more conclusive comparison a
detailed treatment of the freeze-out dynamics is mandatory.  In the
vicinity of $T_c$, the $\Delta$ width increases substantially.  We expect
this trend to be further magnified when including transverse parts in the
vertex corrections, especially in combination with in-medium
$\rho$-mesons~\cite{rw99}.

In the right panel of Fig.~\ref{fig_del} we show the $\Delta$-spectral
function in a net-baryon rich medium, representative for the future GSI
facility. Whereas in dilute matter the line shape is only little affected,
the resonance structure has essentially melted close to $T_c$, mostly due
to a strong renormalization of the pion propagator at high density.

\section{Conclusions and outlook}
\label{sect.5}
Based on hadronic interaction Lagrangians employed within a
finite-temperature many-body approach we have evaluated medium effects on
pions, nucleons and deltas.  The resulting $\Delta$-spectral functions in
cold nuclear matter provide fair agreement with photoabsorption data on
nuclei.  In hot hadronic matter, we found a significant broadening and a
slight upward peak shift of the $\Delta$ resonance, qualitatively in line
with preliminary measurements of $\pi N$ invariant-mass spectra at RHIC.

Future improvements of the $\pi N \Delta$ system in vacuum include
$u$-channel exchange diagrams as well as spin-3/2-$\Delta^*$
excitations which we expect to increase the rather low form-factor
cut-off used so far. 

We further plan to implement in-medium baryon propagators into the 
description of axial-/vector mesons within a chiral framework to 
arrive at a more consistent picture of the equation of state
of hadronic matter under extreme conditions~\cite{vosk04} and the chiral
phase transition.  Another interesting ramification~\cite{Rapp04} 
concerns the role of the medium-modified $\Delta$ spectral functions 
in the soft photon enhancement as recently observed at the 
SPS~\cite{agg03}.

\ack\vspace*{-0.5mm}
One of us (HvH) acknowledges support from the Alexander-von-Humboldt 
Foundation as a Feodor-Lynen Fellow.


\end{document}